\documentclass[conference]{IEEEtran}

\usepackage{amssymb,algorithm,algpseudocode,amsmath,amsfonts,bm,epsfig,graphicx,theorem,epstopdf,wasysym,tabularx}
\usepackage{rotating,setspace,latexsym,epsf,color,dsfont,caption}
\usepackage{cite,subfigure}

\newtheorem{Theorem}{Theorem}

\newtheorem{Lemma}{Lemma}

\newtheorem{Definition}{Definition}

\newenvironment{Proof}[1]{\medskip\par\noindent
{\bf Proof:\,}\,#1}{{\mbox{\,$\blacksquare$}\par}}

\newcommand{\xv}{\mathbf{x}}

\newcommand{\sv}{\mathbf{s}}

\newcommand{\Eb}{\mathbb{E}}
\newcommand{\Pb}{\mathbb{P}}

\newcommand{\lv}{\mathbf{1}}

\newcommand{\tauv}{\boldsymbol{\tau}}

\newcommand{\gammav}{\boldsymbol{\gamma}}

\newcommand{\Sc}{\mathcal{S}}

\newcommand{\Fc}{\mathcal{F}}

\newcommand{\Hc}{\mathcal{H}}

\ifodd 0
\newcommand{\yj}[1]{{\color{blue}#1}}
\else
\newcommand{\yj}[1]{#1}
\fi

\ifodd 0
\newcommand{\yjc}[1]{{\color{red}(YJ: #1)}}
\else
\newcommand{\yjc}[1]{}
\fi

\title{To Skip or to Switch? Minimizing Age of Information under Link Capacity Constraint}

\author{
 \IEEEauthorblockN{Boyu Wang\qquad Songtao Feng \qquad Jing Yang}
    \IEEEauthorblockA{School of Electrical Engineering and Computer Science \\
The Pennsylvania State University\\
 University Park, PA 16802\\
    \emph{\{bxw91,sxf302,yangjing\}@psu.edu}}

\thanks{This work was supported in part by the US National Science Foundation (NSF) under Grant ECCS-1650299.}
}

\begin{document}
\IEEEoverridecommandlockouts
\maketitle
\thispagestyle{empty}

\begin{abstract}
Consider a scenario where a source continuously monitors an object and sends time-stamped status updates to a destination through a rate-limited link. In order to measure the ``freshness" of the status information available at the destination, we adopt the metric called Age of Information (AoI). We assume all updates are of the same size, and arrive randomly at the source according to a Bernoulli process. Due to the link capacity constraint, it takes $d$ ($d\geq 2$) time slots for the source to complete the transmission of an update. Therefore, when a new update arrives at the source during the transmission of another update, the source needs to decide whether to skip the new arrival or to switch to it, in order to minimize the expected average AoI at the destination. We prove that within a broadly defined class of online policies, the optimal policy should be a renewal policy, and has a sequential switching property. We then show that the optimal decision of the source in any time slot has a multiple-threshold structure, and only depends on the age of the update being transmitted and the AoI in the system. The thresholds are then numerically identified by formulating the problem as a Markov Decision Process (MDP). 
\end{abstract}

\section{Introduction}
Enabled by the proliferation of ubiquitous sensing devices and the pervasive wireless data connectivity, real-time monitoring has become a reality in large-scale cyber-physical systems, such as power grids, manufacturing facilities, and smart transportation systems. However, the unprecedented high-dimensionality and generation rate of the sensing data also impose critical challenges on its timely delivery. In order to measure and ensure the freshness of information available to the central controller, a metric called Age of Information (AoI) has been introduced and analyzed in various networks~\cite{infocom/KaulYG12}. Specifically, at time $t$, the AoI in the system is defined as $t-u(t)$, where $u(t)$ is the time stamp of the latest received update at the destination. Since AoI depends on data generation as well as queueing and transmission, it exhibits fundamental differences between traditional network performance metrics, such as throughput and delay.

Modeling the status updating process as a queueing process, time average AoI has been analyzed in systems with a single server~\cite{infocom/KaulYG12,ciss/KaulYG12,isit/YatesK12,YatesK16,Pappas:2015:ICC,isit/NajmN16,isit/KamKNWE16,isit/ChenH16}, and multiple servers \cite{isit/KamKE13, isit/KamKE14,tit/KamKNE16}. Peak Age of Information (PAoI) has been introduced and studied in \cite{isit/CostaCE14,tit/CostaCE16,isit/HuangM15}. The optimality properties of a preemptive Last Generated First Served service discipline are identified in \cite{isit/BedewySS16}. 

AoI minimization has also been investigated, either by controlling the generation process of the updates~\cite{infocom/SunUYKS16,SunPU17,isit/Yates15,ita/BacinogluCU15,Yang:2017:AoI,BacinogluU17,Ahmed:2018:ICC,Ahmed:2018:ITA,Uysal:2018:EH,Feng:2018:INFOCOM,Feng:2018:ISIT}, or by scheduling the transmission of updates that have already been generated~\cite{Modiano:2018:BC,Hsu:2017:ISIT,Hsu:2018:ISIT,Kadota:2018:INFOCOM,He:TIT:2017}. Optimal status updating policy with knowledge of the server state has been studied in~\cite{infocom/SunUYKS16}. AoI-optimal sampling of a Wiener process is investigated in \cite{SunPU17}. Under an energy harvesting setting, optimal status updating have been studied in~\cite{isit/Yates15,ita/BacinogluCU15,Yang:2017:AoI,BacinogluU17,Ahmed:2018:ICC,Ahmed:2018:ITA,Uysal:2018:EH,Feng:2018:INFOCOM,Feng:2018:ISIT}. Transmission scheduling in a broadcast channel has been studied in \cite{Modiano:2018:BC,Hsu:2017:ISIT,Hsu:2018:ISIT}. Reference \cite{Modiano:2018:BC} shows that a greedy policy which always tries to update the most outdated client is optimal in a symmetric setting. Reference \cite{Hsu:2017:ISIT} formulates the problem as a Markov Decision Process (MDP), and show that the optimal policy is a switch-type. It also proposes a sequence of finite-state approximations for the infinite-state MDP and proves its convergence. 
A restless bandits based formulation and a Whittle's index based scheduling have been studied in \cite{Hsu:2018:ISIT}. Different transmission scheduling policies for AoI minimization in a multiple access channel under throughput constraints on individual nodes have been analyzed in \cite{Kadota:2018:INFOCOM}.
Age-optimal link scheduling in a multiple-source system with conflicting links is studied in~\cite{He:TIT:2017}, and the problem is shown to be NP-complete in general.
Head-of-line age-based scheduling algorithms have been shown to be throughput optimal in wireless networks in~\cite{Srikant:2015:ageSchedule}.

In this paper, we investigate the optimal online transmission scheduling for a single link under the assumption that the link capacity is limited and each update takes multiple time slots to transmit. During the transmission of an update, new updates may arrive. Therefore, the source has to decide whether to switch to the new arrival, or to continue its current transmission and drop the new update. What makes the problem challenging is that the impact of a decision on the AoI evolution won't become clear immediately. This is because the instantaneous AoI at the destination will be reset only when a transmission is completed. Even if the source decides to transmit an update at an earlier time, it may drop the update later before the transmission is complete, leading to uncertain AoI evolution in the system. To overcome this challenge, we first prove that within a broadly defined class of online policies, the optimal policy should be a renewal policy, and the decision-making over each renewal interval only depends on the arrival time of the updates in that interval. Then, we show that the optimal renewal policy has a multiple-threshold structure, which enables us to formulate the problem as an MDP, and identify the thresholds numerically through structured value iteration.

\section{System Model and Problem Formulation} \label{sec:model}
We consider a single-link status monitoring system where the source keeps sending time-stamped status updates to a destination through a rate-limited link. We assume the time axis is discretized into time slots, which are labeled as $t=1, 2, 3,\cdots$. At the beginning of time slot $t$, an update packet is generated and arrives at the source according to an independent and identically distributed (i.i.d.) Bernoulli process $A(t)$ with parameter $p$. We assume each update is of the same size, and it takes exactly $d$ time slots, $d\geq 2$, to transmit one update to the destination. Similar to \cite{Modiano:2018:BC,Hsu:2017:ISIT,Hsu:2018:ISIT}, we assume that at most one update can be transmitted during each time slot, and there is {\it no buffer} at the source to store the updates that are not being transmitted. Therefore, once an update arrives at the source, it needs to decide whether to transmit it and drop the one currently under transmission if there is any, or to drop the new arrival.

A status update policy is denoted as $\pi$, which consists of a sequences of transmission decisions $\{D(t)\}$. We let $D(t)\in\{0,1\}$. Specifically, when $A(t)=1$, $D(t)$ can take both values 1 and 0: If $D(t)=1$, the source will start transmitting the new arrival in time slot $t$ and drop the unfinished update if necessary. We term this as {\it switch}; Otherwise, if $D(t)=0$, the source will drop the new arrival, and continue the unfinished transmission. We term this as {\it skip}. When $A(t)=0$, we can show that dropping the update being transmitted is sub-optimal. Thus, we restrict to the policies under which $D(t)$ can only take value 0, i.e., to continue transmitting the unfinished update if there is one, or to idle.

\begin{figure}[t]
	\centering
\includegraphics[width=6.5cm]{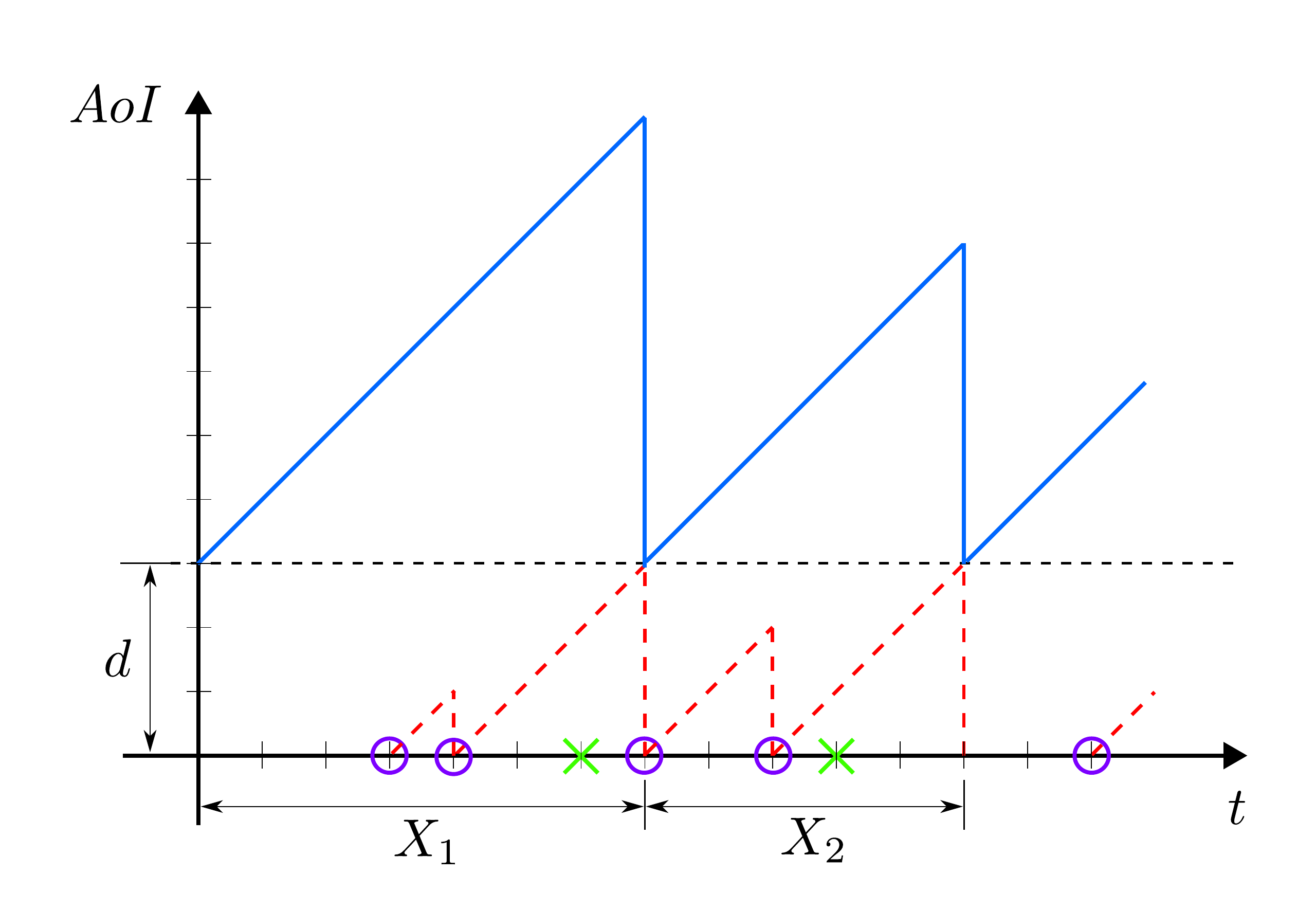}
\vspace{-0.1in}
	\caption{\small{AoI evolution with $d=3$. Circles represent transmitted updates, and crosses represent skipped ones. Red dashed curve indicates the transmitted portion of the corresponding update.} }
	\label{fig:AoI}
	\vspace{-0.1in}
\end{figure}

Let $S_n$ be the the time slot when an update is completely transmitted to the destination. Then,
the inter-update delays can be denoted as $X_n:=S_n-S_{n-1}$, for $n=1,2,\ldots$. Without loss of generality, we assume $S_0=0$. Note that under the bufferless assumption, the AoI after a completed transmission is always equal to $d$. An example sample path of the AoI evolution under a given status update policy is shown in Fig.~\ref{fig:AoI}. 
As illustrated, some updates are skipped when they arrive, while others are transmitted partially or completely.

We use $N(T)$ to denote the total number of successfully delivered status updates over $(0,T]$. 
Define $R(T)$ as the total age of information experienced by the system over $[0,T]$. Denote $R_n:=(2d+X_n)X_n/2$, i.e., the total AoI experienced by the receiver over the $n$th epoch $X_n$. Then,
\begin{align*}
R(T)&=\sum_{n=1}^{N(T)}R_n+ \frac{1}{2}(d+T-S_{N(T)})(T-S_{N(T)}).
\end{align*}
We focus on a set of {\it online} policies $\Pi$, in which the information available for determining $D(t)$ includes the decision history $\{D(i)\}_{i=1}^{t-1}$, the update arrival profile $\{A(i)\}_{i=1}^{t}$, as well as the update statistics (i.e., $p$ in this scenario).
The optimization problem can be formulated as
\begin{eqnarray}\label{eqn:opt}
\underset{\pi\in\Pi}{\min} & &\limsup_{T\rightarrow \infty} \Eb\left[\frac{R(T)}{T}\right]\end{eqnarray}
where the expectation in the objective function is taken over all possible update arrival sample paths.

\section{Structure of the Optimal Policy}
Consider the $n$th epoch, i.e., the duration between time slots $S_{n-1}+1$ and $S_n$ under any online policy in $\Pi$. Let $a_{n,k}$ be the time slot when the $k$th update after $S_{n-1}$ arrives, and let $x_{n,k}:=a_{n,k}-S_{n-1}$. Denote the update arrival profile in epoch $n$ as $\xv_n:=(x_{n,1},x_{n,2},\ldots)$. Then, we introduce the following definition.

\begin{Definition}[Uniformly Bounded Policy]
Under an online policy $\pi\in\Pi$, if there exists a function $g(\xv)$, such that for any $\xv_n=\xv$, the length of the corresponding epoch $X_n$ is upper bounded by $g(\xv)$, and $\Eb[g^2(\xv)]<\infty$, then this policy is a uniformly bounded policy.
\end{Definition}

Denote the subset of uniformly bounded policies as $\Pi'$. Then, using techniques similar to the proof of Theorem 1 in \cite{Ahmed:2018:ICC}, we can show the following theorem.
\begin{Theorem}\label{thm:renewal}
Any uniformly bounded policy $\pi\in\Pi'$ is sub-optimal to a renewal policy. That is, $\{S_n\}_{n=1}^{\infty}$ form a renewal process. Besides, the decision $D(t)$ over the $n$th renewal epoch only depends on $\xv_n$ causally.
\end{Theorem}

Due to space limitation, the proof of Theorem~\ref{thm:renewal}, as well as the proofs of Lemma~\ref{lemma:idle}, Lemma~\ref{lemma:threshold} and Theorem~\ref{thm:threshold} are omitted.

\if{0}
\begin{Proof}
Assume $\{X_i\}$ satisfies uniformly bounded condition. 
For a fixed history $\Hc^{i-1}$, let us group all the status updating sample paths that have the same $\tauv_i$ and perform a statistical averaging over all of them to get the following average age in the $i$th epoch
\begin{align}
\hat{R}_i (\gammav, \Hc^{i-1}) := \Eb [R_i |\tauv_i=\gammav, \Hc^{i-1}]
\end{align}
Then, we have
\begin{align}
\Eb[\hat{R}_i \lv_{i\leq N(T)+1}]=\Eb_{\Hc^{i-1}}[\Eb_{\gammav}[\hat{R}_i (\gammav, \Hc^{i-1})] \lv_{i\leq N(T)+1}|\Hc^{i-1}]
\end{align}
where the equality follows from the fact that $\lv_{i\leq N(T)+1}$ is independent of $\tauv_i$ given $\Hc^{i-1}$. 

Similarly, define the average $i$th epoch length as
\begin{align}
\hat{X}_i(\gammav, \Hc^{i-1}):=\Eb [X_i |\tauv_i=\gammav, \Hc^{i-1}]
\end{align}

Note that for any uniformly bounded policy, we have $\lim_{T\rightarrow\infty}\frac{X^2_{N(T)+1}}{T}=0$ \cite{Feng1806:Minimizing}.
Then,
\begin{align}
&\lim_{T\rightarrow\infty}\frac{\Eb[R(T)]}{T}=\lim_{T\rightarrow\infty}\frac{\Eb\left[\sum_{i=1}^{N(T)+1}R_i\right]}{T}  \label{eqn:thm-1}\\
&\geq \frac{\Eb\left[\sum_{i=1}^{N(T)+1}R_i\right]}{2\Eb[S_{N(T)+1}]}=\frac{ \sum_{i=1}^{\infty} \Eb\left[R_i\lv_{i\leq N(T)+1}\right]}{ \sum_{i=1}^{\infty}\Eb[X_i\lv_{i\leq N(T)+1}]}\\
&=\frac{ \sum_{i=1}^{\infty} \Eb_{\Hc^{i-1}}\left[\Eb_{\gammav}\left[\hat{R}_i(\gammav,\Hc^{i-1})\right]\lv_{i\leq N(T)+1}|\Hc^{i-1} \right]}{ \sum_{i=1}^{\infty}\Eb[X_i\lv_{i\leq N(T)+1}]}\\
&=\frac{ \sum_{i=1}^{\infty} \Eb_{\Hc^{i-1}}\left[\Eb_{\gammav} [\hat{X}_i(\gammav, \Hc^{i-1})]\frac{\Eb_{\gammav}\left[\hat{R}_i(\gammav,\Hc^{i-1})\right]\lv_{i\leq N(T)+1}}{\Eb_{\gammav} [\hat{X}_i(\gammav, \Hc^{i-1})]} |\Hc^{i-1}\right]}{\sum_{i=1}^{\infty}\Eb[X_i\lv_{i\leq N(T)+1}]}\\
&\geq\frac{ \sum_{i=1}^{\infty} \Eb_{\Hc^{i-1}}\left[\Eb_{\gammav} [\hat{X}_i(\gammav, \Hc^{i-1})] R^*\lv_{i\leq N(T)+1} |\Hc^{i-1}\right]}{\sum_{i=1}^{\infty}\Eb[X_i\lv_{i\leq N(T)+1}]}\\
&= R^*
\end{align}
where $R^*:=\min_{i,\Hc^{i-1}}\frac{\Eb_{\gammav}\left[\hat{R}_i(\gammav,\Hc^{i-1})\right]}{\Eb_{\gammav} [\hat{X}_i(\gammav, \Hc^{i-1})]}$, which corresponds to the expected average AoI achieved by a renewal policy under which $X_i$ only depends on $\gammav$.
\end{Proof}
\fi

Based on Theorem~\ref{thm:renewal}, in the following, we will focus on renewal policies that depend on $\xv_n$ only. 

\begin{Lemma}\label{lemma:idle}
If the source is idle when an update arrives, it should start transmitting the update immediately.
\end{Lemma}

\if{0}
\begin{Proof}
 If $t+d=S_n$, the file arrives at $t$ should transmit immediately to reach the receiver at $S_n$. If $t+d<S_n$, transmit the file immediately will not change $S_n$ due to the fact that a new arrival file can replace the transmitting one. Besides, the age of the source keeps growing no matter the source is idle or not. Therefore, if a files arrives at the transmitter when it is idle, the transmitter should start transmitting it immediately.
\end{Proof}

We note the the optimal policy is to decide whether the transmitter should switch to a new update when it arrives.
\fi
\begin{Definition}[Sequential Switching Policy]
A sequential switching (SS) policy is a renewal policy under which the source switches to an update arriving at time slot $t$ \textbf{only if} it switches to all update arrivals prior to $t$ in the same epoch. 
\end{Definition}

\textbf{Remark:} The definition of SS policy implies that once a source skips a new update arrival at $t$, it will skip all of the upcoming update arrivals until it finishes the one being transmitted at $t$. We point out that an SS policy is in general different from threshold type of policies, as it does not impose any threshold structure on when the source should skip or switch to a new update arrival.

\begin{Lemma}\label{lemma:ssp}
The optimal renewal policy in $\Pi'$ is an SS policy. 
\end{Lemma}
\begin{Proof}
We prove this lemma through contradiction. Now assume the optimal policy $\pi_0$ is not an SS policy. Without loss of generality, we consider the first renewal epoch starting at time 0 (the beginning of time slot $1$). We assume under $\pi_0$ there exists a sample path under which the source transmits the new update arrival at time slot $i$ and does not switch to the next arrival at time slot $j$ in the same epoch, i.e., $i<j<i+d$. Depending on the upcoming random arrivals, the sample path may evolve into different sample paths. Denote the set of such sample paths as $\Fc_j$, as they share the same history up to time slot $j$. We can partition $\Fc_{j}$ into two subsets:
\begin{itemize}
\item $\Fc_{j,1}$: The source skips all the upcoming arrivals and finishes transmitting the update arrives at $i$.
\item $\Fc_{j,2}$: The source switches to some later arrival.
\end{itemize}
Let $X^{\pi_0}$ be the corresponding length of the renewal epoch under policy $\pi_0$. Then, $X^{\pi_0}=i+d-1$ for sample paths in $\Fc_{j,1}$, and $X^{\pi_0}>j+d-1$ for sample paths in $\Fc_{j,2}$.

We now construct two policies $\pi_1$ and $\pi_2$ as follows. Under both $\pi_1$ and $\pi_2$, the source will behave exactly the same as under $\pi_0$ for all sample paths not in $\Fc_j$. However, for the sample paths in $\Fc_j$, the actions the source will take after $j$ will be different. Specifically, under $\pi_1$, the source will finish the update that arrives at time slot $i$ irrespective of other factors. Therefore, for all sample paths in $\Fc_j$ under $\pi_0$, the corresponding length of the renewal epoch under $\pi_1$ will be $X^{\pi_1}=i+d-1$ under $\pi_1$. For $\pi_2$, we will let the source first switch to the arrival at time slot $j$, and then switch to a later arrival whenever the source switches under $\pi_0$. Then, for the sample paths in $\Fc_{j,1}$ under $\pi_0$, the corresponding length of renewal epoch will be changed to $X^{\pi_2}=j+d-1$ under $\pi_2$; while for those in $\Fc_{j,2}$,  $X^{\pi_2}=X^{\pi_0}$.

Therefore, considering all possible sample paths under those policies, we have
$\Eb[X^{\pi_1}]<\Eb[X^{\pi_0}]<\Eb[X^{\pi_2}]$,
which implies that there must exist a $\rho$, $0<\rho<1$, such that
\begin{align}\label{eqn:E[X]}
\rho \Eb[X^{\pi_1}]+(1-\rho)\Eb[X^{\pi_2}]=\Eb[X^{\pi_0}].
\end{align}

We will then construct a randomized policy $\pi'$, under which it follows $\pi_1$ with probability $\rho$ and follows $\pi_2$ with probability $1-\rho$. Apparently, the expected length of the renewal epoch under $\pi'$, denoted as $X^{\pi'}$, will be the same as that under $\pi_0$. 

Next, we will show that $\Eb[(X^{\pi'})^2]\leq \Eb[(X^{\pi_0})^2]$. Denote $P_1:=\frac{\Pb_{\pi_0}[\Fc_{j,1}]}{\Pb_{\pi_0}[\Fc_j]},P_{2}:=\frac{\Pb_{\pi_0}[\Fc_{j,2}]}{\Pb_{\pi_0}[\Fc_j]}$. Then, (\ref{eqn:E[X]}) can be expressed as
\begin{align}\label{eqn:E[X]_2}
&\rho(i+d-1)+(1-\rho)\left[(j+d-1)P_1+ \Eb[X^{\pi_0}|\Fc_{j,2}]P_{2}\right]\nonumber\\
&=(i+d-1)P_1+\Eb[X^{\pi_0}|\Fc_{j,2}]P_{2},
\end{align} 
which can be reduced to
\begin{align}\label{eqn:E[X]_3}
&(1-\rho)P_{1}(j+d-1)\nonumber\\
&=(P_1-\rho) (i+d-1)+\rho P_{2}\Eb[X^{\pi_0}|\Fc_{j,2}].
\end{align}
Since $\Eb[X^{\pi_0}|\Fc_{j,2}]>j+d-1$, $(1-\rho)P_{1}=(P_1-\rho)+\rho P_{2}$, (\ref{eqn:E[X]_3}) implies that
$P_1-\rho>0$.
Dividing both sides of (\ref{eqn:E[X]_3}) by $(1-\rho)P_{1}$, we have
\begin{align*}
j+d-1&=\frac{P_1-\rho}{(1-\rho)P_{1}}(i+d-1)+\frac{\rho P_{2}}{(1-\rho)P_{1}} \Eb[X^{\pi_0}|\Fc_{j,2}].
\end{align*}
Note that $\frac{P_1-\rho}{(1-\rho)P_{1}}$ and $\frac{\rho P_{2}}{(1-\rho)P_{1}}$ form a valid distribution. Therefore, based on Jensen's inequality, we have
\begin{align}
&(j+d-1)^2\nonumber\\
& < \frac{P_1-\rho}{(1-\rho)P_{1}} (i+d-1)^2+ \frac{\rho P_{2}}{(1-\rho)P_{1}} \left(\Eb[X^{\pi_0}|\Fc_{j,2}]\right)^2\\
&\leq\frac{P_1-\rho}{(1-\rho)P_{1}} (i+d-1)^2+ \frac{\rho P_{2}}{(1-\rho)P_{1}} \Eb[\left(X^{\pi_0}\right)^2|\Fc_{j,2}],
\end{align}
which is equivalently to
\begin{align}\label{eqn:E[X^2]}
&\rho(i+d-1)^2+(1-\rho)\left[(j+d-1)^2P_1+ \Eb[\left(X^{\pi_0}\right)^2|\Fc_{j,2}]P_{2}\right]\nonumber\\
&<(i+d-1)^2P_1+\Eb[(X^{\pi_0})^2|\Fc_{j,2}]P_{2}.
\end{align} 
I.e.,
\begin{align}\label{eqn:E[X^2]_2}
\rho \Eb[(X^{\pi_1})^2]+(1-\rho)\Eb[(X^{\pi_2})^2]<\Eb[(X^{\pi_0})^2].
\end{align}
Combining (\ref{eqn:E[X]}) and (\ref{eqn:E[X^2]_2}), we have 
\begin{align}
\frac{1}{2}\frac{\rho \Eb[(X^{\pi_1})^2]+(1-\rho)\Eb[(X^{\pi_2})^2]}{\rho \Eb[X^{\pi_1}]+(1-\rho)\Eb[X^{\pi_2}]}< \frac{1}{2}\frac{\Eb[(X^{\pi_0})^2]}{\Eb[X^{\pi_0}]},
\end{align}
i.e., the new policy $\pi'$ achieves a lower expected average AoI than $\pi_0$, which contradicts with the assumption that $\pi_0$ is optimal.
\end{Proof}

\begin{Lemma}\label{lemma:threshold}
Under the optimal SS policy in $\Pi'$, if the source is transmitting an update that arrives at the $i$th time slot in an epoch when the new update arrives, then, there exists a threshold $\tau_i$, $\tau_i\geq i$, which depends on $i$ only, such that if the new update arrives before or at the $\tau_i$th time slot in that epoch, the source will switch to the new arrival; otherwise, it will skip the new arrival and complete the current transmission.
\end{Lemma}

\if{0}
\begin{Proof}
We prove this lemma through contradiction. Again, we focus on the first renewal epoch that starts at time 0. Assume under the optimal SS policy $\pi_0$, the transmitter transmits the update that arrives in time slot $i$. It will then skip the next arrival if it arrives at time slot $j$, and switches to it if it arrives at time slot $k$, with $j<k<i+d$. We aim to perturb $\pi_0$ a bit to obtain a new policy $\pi'$ and show that it achieves a lower expected average AoI. 

Similarly to the proof of Lemma~\ref{lemma:ssp}, we consider all sample paths under $\pi_0$ that choose to transmit the update that arrives at time slot $i$, while the next arrival time is either $j$ or $k$. Denote the subset of such sample paths as $\Fc_i$. Again, we can divide $\Fc_i$ into two subsets.
\begin{itemize}
\item $\Fc_{i,1}$: The next arrival time is $j$. Then, the sample paths will skip $j$ and finish transmitting $i$. The corresponding length of renewal interval $X^{\pi_0}=i+d-1$.
\item $\Fc_{i,2}$: The next arrival time is $k$. The sample paths will switch to the new arrival, and the corresponding length of renewal interval $X^{\pi_0}\geq k+d-1$.
\end{itemize}

Then, we construct the new policy $\pi'$ in this way: For all sample paths in $\Fc_{i,1}$, the new policy will let the transmitter randomly switch to the next arrival at $j$ and finish transmitting it with probability $P_j$, and let the rest sample paths follow the original policy $\pi_0$, i.e., skip the new arrival and finish transmitting the arrival at $i$. For all sample paths in $\Fc_{a,2}$, the new policy will let the transmitter randomly skip the next arrival at $k$ with probability $P_k$, and let the rest sample paths follow the original policy $\pi_0$. Then, under $\pi'$, we have
\begin{align}
\Eb[X^{\pi'}|\Fc_{i,1}]&= P_j (j+d-1) +(1-P_j)(i+d-1)\\
\Eb[X^{\pi'}|\Fc_{i,2}]&= P_k (i+d-1) +(1-P_k)\Eb[X^{\pi_0}|\Fc_{i,2}]
\end{align}
We pick $P_j$ and $P_k$ in a way such that
\begin{align}
\Eb[X^{\pi'}|\Fc_i]&=\Eb[X^{\pi_0}|\Fc_i]  \label{eqn:lemma3-11}
\end{align}
i.e.,
\begin{align*}
P_j (j-i) \Pb[\Fc_{i,1}] &=P_k [\Eb[X^{\pi_0}|\Fc_{i,2}]- (i+d-1)] \Pb[\Fc_{i,2}]   
\end{align*}
Following similar argument as in the proof of Lemma~\ref{lemma:ssp}, we can show that
\begin{align}
\Eb[(X^{\pi'})^2|\Fc_i]&\leq\Eb[(X^{\pi_0})^2|\Fc_i]   \label{eqn:lemma3-0}
\end{align}

\if{0}
Separating term $(b+d)$ from equation (\ref{eqn:lemma3-1}), we have
\begin{align}
b+d=\frac{P_b \Pb[\Fc^I_a] \hspace{-0.03in} - \hspace{-0.03in} P_c\Pb[\Fc^{II}_a]}{P_b \Pb[\Fc^I_a]}(a+d)  
+ \frac{P_c\Pb[\Fc^{II}_a]}{P_b \Pb[\Fc^I_a]} \Eb[X^{\pi_0}|\Fc^{II}_a]
\end{align}
Since $\Eb[X^{\pi_0}|\Fc^{II}_a]\geq c+d$ and $a<b<c$, from equation (\ref{eqn:lemma3-1}) we must have
\begin{align}
\frac{P_c\Pb[\Fc^{II}_a]}{P_b \Pb[\Fc^I_a]} = \frac{b-a}{\Eb[X^{\pi_0}|\Fc^{II}_a]-(a+d)} \leq \frac{b-a}{c-a} 
\leq 1
\end{align}
Thus $\frac{P_b \Pb[\Fc^I_a]- P_c\Pb[\Fc^{II}_a]}{P_b \Pb[\Fc^I_a]}$ and $\frac{P_c\Pb[\Fc^{II}_a]}{P_b \Pb[\Fc^I_a]}$ form a valid distribution.

From Jensen's inequality, we have
\begin{align}
(b+d)^2 &\leq \frac{P_b \Pb[\Fc^I_a]\hspace{-0.03in} -\hspace{-0.03in} P_c\Pb[\Fc^{II}_a]}{P_b \Pb[\Fc^I_a]}(a+d)^2 \nonumber \\
&\qquad \qquad + \frac{P_c\Pb[\Fc^{II}_a]}{P_b \Pb[\Fc^I_a]} \left( \Eb[X^{\pi_0}|\Fc^{II}_a] \right)^2 \\
&\leq \frac{P_b \Pb[\Fc^I_a]\hspace{-0.03in} -\hspace{-0.03in} P_c\Pb[\Fc^{II}_a]}{P_b \Pb[\Fc^I_a]}(a+d)^2 \nonumber \\
&\qquad \qquad + \frac{P_c\Pb[\Fc^{II}_a]}{P_b \Pb[\Fc^I_a]}  \Eb[(X^{\pi_0})^2|\Fc^{II}_a] \label{eqn:lemma3-2}
\end{align}
Plug (\ref{eqn:lemma3-2}) in the left hand side of (\ref{eqn:lemma3-0}), we have
\begin{align}
LHS&=\Pb[\Fc^I_a]\left( P_b(b+d)^2 + (1-P_b) (a+d)^2 \right)  \nonumber \\
&\quad + \Pb[\Fc^{II}_a]\left( P_c (a+d)^2 + (1-P_c) \Eb[(X^{\pi_0})^2|\Fc^{II}_a] \right)  \\
&\leq \Pb[\Fc^I_a] (a+d)^2 + \Pb[\Fc^{II}_a] \Eb[(X^{\pi_0})^2|\Fc^{II}_a] \\
&=RHS
\end{align}
which implies (\ref{eqn:lemma3-0}) holds. 
\fi

Thus policy $\pi'$ achieves lower expected average AoI which follows from (\ref{eqn:lemma3-11}), which contradicts with the assumption that $\pi_0$ is optimal. Therefore, if the source skips the next update if it arrives at time slot $j$, it must skip it if it arrives later than $j$. Picking the minimum $j$ ($j>i$) in which a skip would occur, denote it as $j^*$. Then, the threshold $\tau_i$ equals $j^*-1$. Since the proof does not depend other information except $i$, $\tau_i$ depends on $i$ only. 
\end{Proof}
\fi

\begin{Theorem}\label{thm:threshold}
Under the optimal policy in $\Pi'$, there exists a sequence of thresholds $\tau_1\geq\tau_2\geq \cdots \geq \tau_K$, such that if the source is transmitting an update that arrives in the $i$th ($i\leq K$) time slot in an renewal epoch when a new update arrives, and the arrival time of the new update is before or at the $\tau_i$th time slot in the epoch, the source will switch to the new arrival; Otherwise, if the next update arrives after $\tau_i$, or the update being transmitted arrives after $K$, the source will skip all upcoming arrivals until it finishes the current transmission. 
\end{Theorem}

Theorem~\ref{thm:threshold} indicates that the optimal decision of the source only depends on two parameters: the arrival time of the update being transmitted, and the arrival time of the new update, both relative to the beginning of the renewal epoch. Therefore, the problem is essentially an MDP. In Sec.~\ref{sec:mdp}, we will cast the problem as an MDP, and numerically search for the optimal thresholds $\tau_1,\tau_2, \cdots \tau_K$ and $K$.

\if{0}
\begin{Proof}
	For successful transmitting time $d=1$, we should always switch according to Lemma \ref{lemma:idle}. Thus we consider $d>1$ in the following proof.
	We consider two scenarios.
	
	1) $\tau_a>a,\tau_{a+1}>a+1$.
	
	We prove in this case the switch thresholds has monotonicity property, i.e., $\tau_{a+1}\leq \tau_a$, $\forall a\geq 0$. 
	
	Note that when $\tau_{a}=a+1$, the result is obvious. We consider the case when $\tau_{a}>a+1$ which implies $\tau_{a+1}>a+2$.
	
	Assume under optimal policy $\pi_0$ there exists two adjacent arrival times $a$ and $a+1$ such that $\tau_{a+1}>\tau_a>a+1>a$. Consider two subset of sample paths as follows.
	
	\begin{itemize}
		\item[I)] $\Fc_{a,\tau_a+1}$: The transmitter transmits the arrival at $a$ and the next arrival is at $\tau_a+1$. Then, under $\pi_0$, it will skip the arrival and finish transmitting the arrival at $a$. The corresponding $X^{\pi_0}=a+d$. 
		\item[II)] $\Fc_{a+1,\tau_a+1}$: The transmitter transmits the arrival at $a+1$ and the next arrival is at $\tau_a+1$. Since  $\tau_a<\tau_{a+1}$, under $\pi_0$, it will switch to the new arrival. The corresponding $X^{\pi_0}\geq \tau_{a}+1+d$. 
	\end{itemize}
	Then, we will modify $\pi_0$ to $\pi'$ as follows:
	For sample paths in $\Fc_{a,\tau_a+1}$, the transmitter will randomly switch to the new arrival at $\tau_a$ with probability $P_a$ and then follow policy $\pi_0$ afterwards. Note that the sample paths after switching at $\tau_a+1$ should be the same as those in $\Fc_{a+1,\tau_a+1}$.  Then, we have
	\begin{align}
	\Eb[X^{\pi'}|\Fc_{a,\tau_a+1}]&=P_a \Eb[X^{\pi_0}|\Fc_{a+1,\tau_a+1}]+(1-P_a) (a+d)
	\end{align}
	For sample paths in $\Fc_{a+1,\tau_a+1}$, the transmitter will randomly skip the new arrival at $\tau_a+1$ with probability $P_{a+1}$ and finish transmitting the arrival at $a+1$; It will stick with policy $\pi_0$ for the rest sample paths. Then
	\begin{align}
	&\Eb[X^{\pi'}|\Fc_{a+1,\tau_a+1}]=P_{a+1} (a+1+d)  \nonumber \\
	&\qquad \qquad\qquad \qquad\quad+(1-P_{a+1}) \Eb[X^{\pi_0}|\Fc_{a+1,\tau_a+1}]
	\end{align}  
	We pick $P_a$ and $P_{a+1}$ in a way such that
	\begin{align}\label{eqn:mean}
	&\Eb[X^{\pi'}|\Fc_{a,\tau_a+1}]P_I+\Eb[X^{\pi'}|\Fc_{a+1,\tau_a+1}]P_{II}\\
	&=\Eb[X^{\pi_0}|\Fc_{a,\tau_a+1}]P_I+\Eb[X^{\pi_0}|\Fc_{a+1,\tau_a+1}]P_{II}
	\end{align}
	where we define $P_I:=\frac{\Pb[\Fc_{a,\tau_a+1}]}{\Pb[\Fc_{a,\tau_a+1}]+\Pb[\Fc_{a+1,\tau_a+1}]}$, $P_{II}:=\frac{\Pb[\Fc_{a+1,\tau_a+1}]}{\Pb[\Fc_{a,\tau_a+1}]+\Pb[\Fc_{a+1,\tau_a+1}]}$.
	Equation (\ref{eqn:mean}) implies that
	\begin{align}
	&P_{a+1}P_{II}(a+1+d) =(P_{a+1}P_{II}-P_aP_I)\Eb[X^{\pi_0}|\Fc_{a+1,\tau_a+1}] \nonumber \\
	&\qquad \qquad\qquad \qquad\qquad \qquad+ P_a P_I (a+d) \label{eqn:lemma4-1}
	\end{align}
	Since $a+d<b+d<\Eb[X^{\pi_0}|\Fc_{a+1,\tau_a+1}]$, from equation (\ref{eqn:lemma4-1}) we must have
	\begin{align}
	\frac{P_aP_I}{P_{a+1}P_{II}}=\frac{\Eb[X^{\pi_0}|\Fc_{a+1,\tau_a+1}]-(a+1-d)}{\Eb[X^{\pi_0}|\Fc_{a+1,\tau_a+1}]-(a+d)}<1
	\end{align}
	Thus $\frac{P_{a+1}P_{II}-P_aP_I}{P_{a+1}P_{II}}$ and $\frac{P_aP_I}{P_{a+1}P_{II}}$ form a valid distribution.
	Following similar steps as in the proof of Lemma~\ref{lemma:ssp}, we have
	\begin{align}
	&P_{a+1}P_{II}(b+d)^2 \leq(P_{a+1}P_{II}-P_aP_I)\Eb[(X^{\pi_0})^2|\Fc_{a+1,\tau_a+1}] \nonumber \\
	& \qquad\qquad\qquad\qquad\qquad+ P_a P_I (a+d)^2
	\end{align}
	which implies that
	\begin{align}\label{eqn:second_moment}
	\Eb[(X^{\pi'})^2\leq\Eb[(X^{\pi_0})^2]
	\end{align}
	Thus, $\pi_0$ cannot be optimal.
	
	2) $\tau_{a}=a$.
	
	We prove in this case $\tau_{a+1}=a+1$.
	
	Assume under optimal policy $\pi_0$ there exists $\tau_{a+1}>a+1$. Consider two subset of sample paths as follows.
	\begin{itemize}
		\item[I)] $\Fc_{a,a+2}$: The transmitter transmits the arrival at $a$ and the next arrival is at $a+2$. Then, under $\pi_0$, it will skip the arrival and finish transmitting the arrival at $a$. The corresponding $X^{\pi_0}=a+d$. 
		\item[II)] $\Fc_{a+1,a+2}$: The transmitter transmits the arrival at $a+1$ and the next arrival is at $a+2$. Since $\tau_{a+1}>a+1$, under $\pi_0$, it will switch to the new arrival. The corresponding $X^{\pi_0}\geq a+2+d$.
	\end{itemize}
	Then, we modify $\pi_0$ to $\pi'$ as follows. For sample paths in $\Fc_{a,a+2}$, the transmitter will randomly switch to the new arrival at $a+2$ with probability $P_a$ and then follow policy $\pi_0$ afterwards. Note that the sample paths after switching at $a+2$ should be the same as those in $\Fc_{a+1,a+2}$.
	
	For sample paths in $\Fc_{a+1,a+2}$, the transmitter will randomly skip the new arrival at $a+2$ with probability $P_{a+1}$ and finish transmitting the arrival at $a+1$; Otherwise it will follow policy $\pi_0$ for the rest sample paths.

	Apply the same technique in 1). Pick $P_a$ and $P_{a+1}$ such that $E[X]$ does not change but $\Eb[(X^{\pi'})^2\leq\Eb[(X^{\pi_0})^2]$. Thus $\pi_0$ is not optimal.
	
	From 1) and 2) we conclude lemma \ref{lemma:threshold} holds.
\end{Proof}
\fi

\section{MDP based Scheduling}\label{sec:mdp}
\subsection{MDP formulation}
Motivated by the Markovian structure of the optimal policy in Theorem~\ref{thm:threshold}, we formulate the problem as an MDP as follows.

\textbf{States:} We define the state $S(t):=(\Delta(t), L(t), A(t))$, where $\Delta(t)$ and $L(t)$ are the AoI in the system, and the age of the unfinished update, at the beginning of time slot $t$, respectively. $A(t)$ is the update arrival status. \yj{Then, $\Delta(t)\geq d$, $0\leq L(t)\leq d-1$, and the state space $\Sc$ can be determined accordingly.}

\textbf{Actions:} $D(t)\in\{0,1\}$, as defined in Sec.~\ref{sec:model}. 

\textbf{Transition probabilities:} The transition probability from a state $\sv:=(\delta, l, \lambda)$ to another state $\sv'$ under action $a$, denoted as $P_{\sv\sv'}(a)$, is shown in Table \ref{Tab:tranP}.

\textbf{Cost:} Let $C(S(t), D(t))$ be the immediate cost after the action $D(t)$ is taken at $t$ under state $S(t)$. We consider the instantaneous AoI after the action as the immediate cost, i.e.,
\begin{align*}
 C(S(t), D(t))&=\left\{ \begin{array}{ll}
 d& \mbox{if $L(t)=d-1, D(t)=0$};\\
 \Delta(t)+1& \mbox{otherwise}.\\
 \end{array}
 \right. 
\end{align*}
 
 

In order to reduce the computational complexity, we define an approximate MDP as follows: We define $\delta_m$ as the boundary AoI, and truncate the state space of the original MDP as $\Sc_m=\{\sv\in \Sc: \delta\leq \delta_m\}$. In the transition probabilities, we bound $\delta+1$ by $\delta_m$, i.e., $[\delta+1]^+_m=\min{(\delta+1,\delta_m)}$.
 
Then, the optimal policy can be determined through relative value iteration as follows:
 \begin{align}\label{eqn:value iteration} 
V_{n+1}(\sv)\hspace{-0.03in}=\hspace{-0.03in}\min_{a\in\{0, 1\}} \hspace{-0.03in} C(\sv, a)\hspace{-0.03in}+\hspace{-0.03in}\sum_{s'} P_{\sv\sv'}(a)V_n(\sv')\yj{-V_n(\sv_0)},
 \end{align}
where $\sv_0$ is a reference state and we set it as $\sv_0:=(d,0,0)$. For each iteration $n$, we need to update the optimal cost function for all states $\sv\in \Sc_m$ by minimizing the right hand side of (\ref{eqn:value iteration}), which causes a high computational complexity as the number of states increases. Motivated by \cite{Hsu:2017:ISIT}, we then leverage the multi-threshold structure of the optimal policy to reduce the computational complexity, as detailed in the structured value iteration algorithm in Algorithm \ref{algorithm:ssa}.

With the multiple-threshold structure, Algorithm \ref{algorithm:ssa} does not need to seek the optimal action by equation (\ref{eqn:value iteration}) for all states in each iteration as the traditional value iteration algorithm does. Specifically, if the optimal action for a state $(\delta',l,1)$ is to {\it skip} the new arrival, the optimal action for state $(\delta, l ,1)$, $\delta>\delta'$ must be to skip as well. Similarly, if the optimal action for a state $(\delta,l',1)$ is to switch to the new arrival, the optimal action for state $(\delta, l ,1)$, $l<l'$, must be to switch. 


 \begin{table}[t]
 \vspace{0.05 in}
	\centering
	\resizebox{9cm}{!}{\begin{tabular}{c|c|c}
		\hline\hline
		$P_{\sv\sv'}(a)$& $a=0$ & $a=1$\\
		\hline
		$l=0$& $\begin{aligned}
			P[(\delta+1, 0, 1)|(\delta, l, \lambda)]&= p \\
			P[(\delta+1, 0, 0)|(\delta, l, \lambda)]&= 1-p \end{aligned}$
		& $\begin{aligned}
			P[(\delta+1, 0, 1)|(\delta, l, 0)] &= p\\
			P[(\delta+1, 0, 0)|(\delta, l, 0)] &= 1-p\\
			P[(\delta+1, 1, 1)|(\delta, l, 1)] &= p\\
			P[(\delta+1, 1, 0)|(\delta, l, 1)] &= 1-p\end{aligned}$\\
			\hline
				$0<l<d-1$& $\begin{aligned}
					P[(\delta+1, 0, 1)|(\delta, l, \lambda)]&= p \\
					P[(\delta+1, 0, 0)|(\delta, l, \lambda)]&= 1-p \end{aligned}$
				& $\begin{aligned}
					P[(\delta+1, 1, 1)|(\delta, l, 1)] &= p\\
					P[(\delta+1, 1, 0)|(\delta, l, 1)] &= 1-p\\
						P[(\delta+1, l+1, 1)|(\delta, l, 0)] &= p\\
						P[(\delta+1, l+1, 0)|(\delta, l, 0)] &= 1-p\end{aligned}$\\
				\hline
					$l=d-1$& $\begin{aligned}
						P[(d, 0, 1)|(\delta, l, \lambda)] &= p \\
						P[(d, 0, 0)|(\delta, l, \lambda)]&= 1-p \end{aligned}$ 
					& $\begin{aligned}
						P[(d, 0, 1)|(\delta, l, 0)] &= p\\
						P[(d, 0, 0)|(\delta, l, 0)] &= 1-p\\
						P[(\delta+1, 1, 1)|(\delta, l, 1)] &= p\\
						P[(\delta+1, 1, 0)|(\delta, l, 1)] &= 1-p\end{aligned}$\\
		\hline\hline
	\end{tabular}}
	\caption{Transition probabilities.}\label{Tab:tranP}
\end{table}

 \begin{algorithm}[t]
 	\caption{Structured Value Iteration.}\label{algorithm:ssa}
 	\begin{algorithmic}[1]
 		\State Initialize: $V_0(\sv)=0,\forall \sv\in \Sc_m$.
 		\For{$i=0 : n$}
 		\For{$\forall \sv\in \Sc_m$}
 		\If{$\lambda=0$}
 		\State $a^*(\sv)=0$;
 		\ElsIf{$\exists \delta'<\delta, a^*(\delta', l, 1)=0$}
 		\State{$a^*(\sv)=0$};
 		\ElsIf{$\exists l'\yj{>}l, a^*(\delta, l', 1)=1$}
 		\State{$a^*(\sv)=1$};
 		\Else
 		\State $a^*(\sv)\hspace{-0.05in}=\hspace{-0.05in}\arg\min_{a\in\{0, 1\}}\hspace{-0.04in}C(\sv,a)\hspace{-0.04in}+\hspace{-0.04in}\sum_{\sv'}\hspace{-0.04in} P_{\sv\sv'}\hspace{-0.02in}(\hspace{-0.02in}a\hspace{-0.02in})\hspace{-0.02in}V_i(\sv')$
 		\EndIf
 		\State $V_{i+1}(\sv)\hspace{-0.03in}=\hspace{-0.03in}C(\sv, a^*(\sv))+\sum_{\sv'} P_{\sv\sv'}(a)V_i(s')-V_i(\sv_0)$
        \EndFor 
 		\EndFor
 		\State \textbf{return} $a^*(\sv), V(\sv).$
 	\end{algorithmic}
 \end{algorithm}
\subsection{Numerical results}
\begin{figure}[t]
	\centering
	\begin{minipage}[t]{4.3cm}
		\centering
		\centerline{\includegraphics[width=4.2cm]{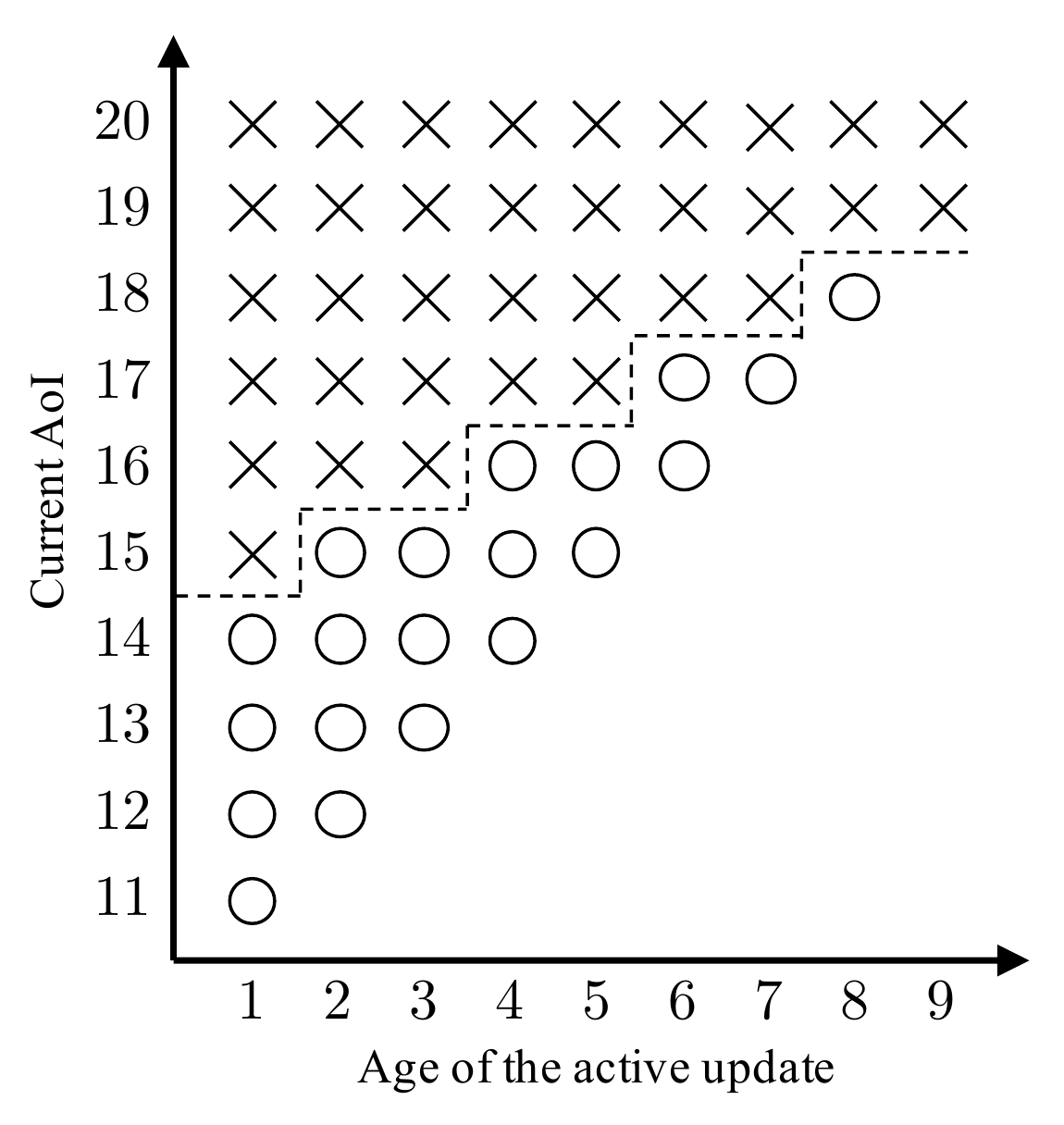}}
		\vspace{-0.05in}
		\centerline{\small{(a)}}
		\label{fig:mdp}
		
	\end{minipage}
	\begin{minipage}[t]{4.2cm}
		\centering
		\centerline{\includegraphics[width=4.3cm]{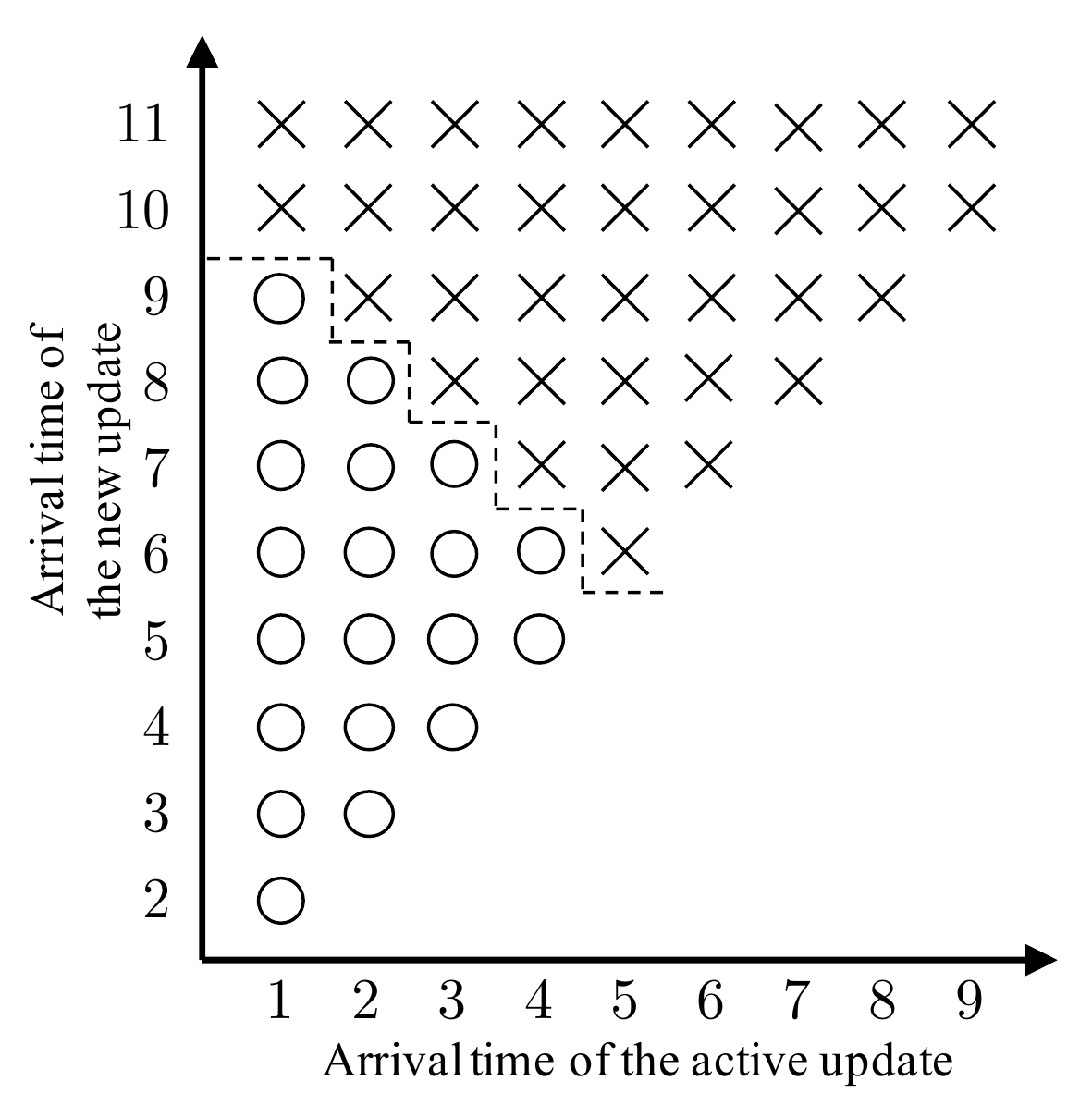}}
				\vspace{-0.05in}
				\centerline{\small{(b)}}
		\label{fig:amdp}
	\end{minipage}
	\caption{The optimal policy when $p=0.07$, $d=10$. Circles represent {\it switch}, while crosses represent {\it skip}.}
	\label{fig:switch}
			\vspace{-0.1in}
\end{figure}

We then search for the optimal policy numerically using Algorithm~\ref{algorithm:ssa}. 
We set $d=10$, $p=0.07$, and \yj{the number of iterations} to be $10,000$.\yjc{?} We set $\delta_m=50$ \yjc{too small?} for the approximate MDP. Fig.~\ref{fig:switch}(a) shows the optimal action for each state $(\Delta(t),L(t),1)$. We then plot the optimal action for each pair of arrival time of the update being transmitted and that of the new arrival in a renewal epoch in Fig.~\ref{fig:switch}(b). We note the thresholds {$\tau_1=9$, $\tau_2=8$, $\tau_3=7$, $\tau_4=6$}. They are monotonically decreasing, as predicted by Theorem~\ref{thm:threshold}. When the update being transmitted arrives after the fourth time slot in that epoch, all upcoming updates will be skipped. 

\begin{figure}[t]
	\centering
	\includegraphics[width=7cm]{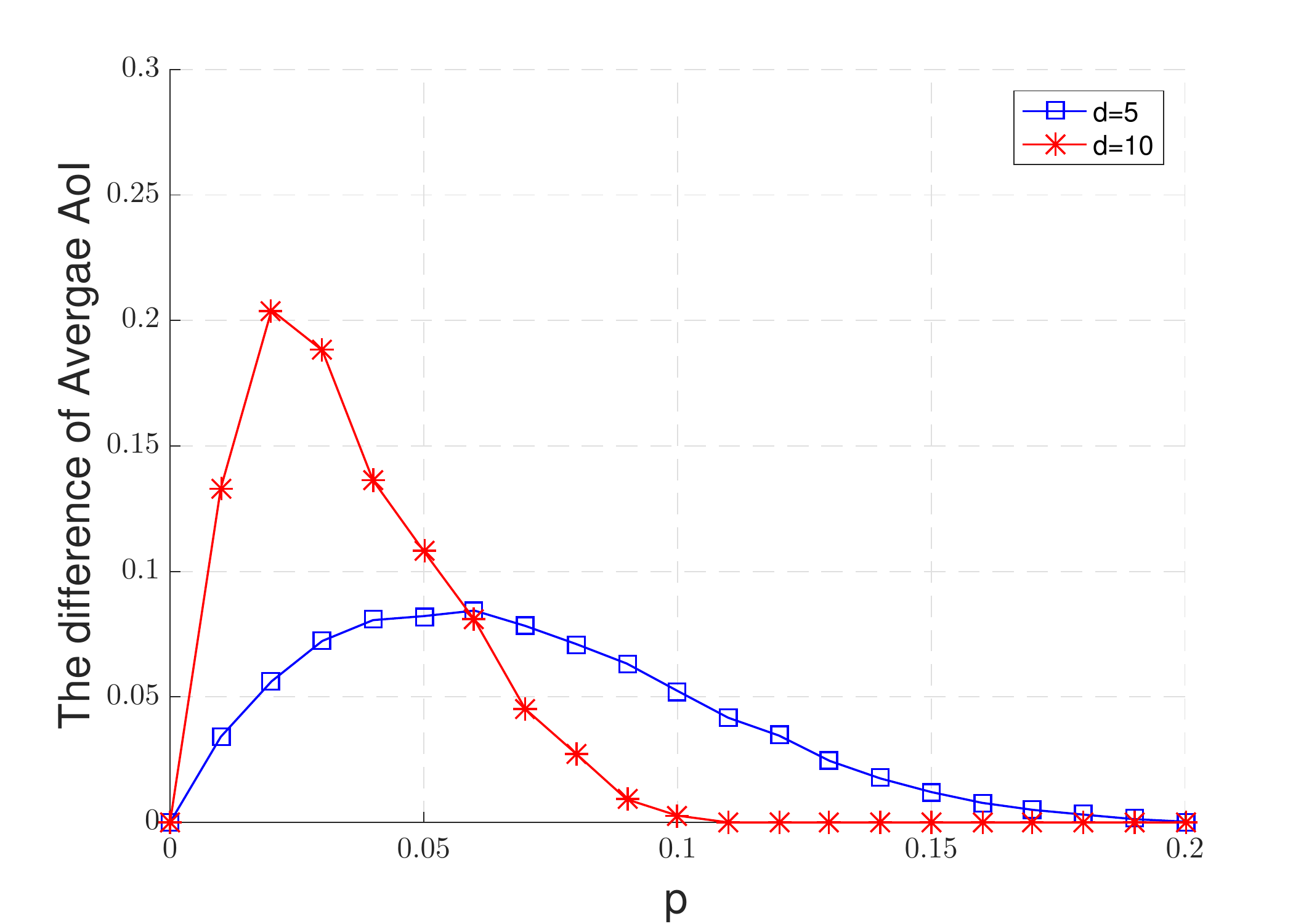}
		\vspace{-0.1in}
	\caption{Performance gap between the optimal multiple-threshold policy and a myopic policy.}
	\label{fig:ageResults2}
	\vspace{-0.1in}
\end{figure}

Then, we compare the average AoI under the optimal policy identified by Algorithm~\ref{algorithm:ssa} and a myopic policy over $10,000$ time slots. Under the myopic policy, the source will never switch to a new update arrival until it finishes the one being transmitted. The performance gap is plotted in Fig. \ref{fig:ageResults2}. As we observe, the optimal policy always outperforms the myopic policy. Although the greedy policy minimizes the length of the each epoch greedily, it does not render the minimum average AoI. This is because $X_i$ has a larger second moment in this case, leading to higher AoI. We note that when $p$ gets sufficiently small or large, the performance gap between both policies becomes close to zero. This is because for such extreme cases, the multiple-threshold policy and the myopic policy become identical to each other.

\if{0}
\begin{figure}[t]
	\centering
	\centerline{\includegraphics[width=8cm]{./graph/result1.eps}}
	\caption{Average age under different transmitting time.}
	\label{fig:ageResults1}
\end{figure}
\fi

\if{0}
Tab. \ref{Tab:runningtime} shows the average running times of the algorithms. We note that the SWS has significant improvement over the traditional value iteration, as the approximate MDP reduces the state spaces and the reduced iteration according to the sequential switching feature in the SWS.

\begin{table}[h]
	\begin{center}
		\caption{Average running times in seconds.}\label{Tab:runningtime}
		{\begin{tabular}{r||c| c| c}
				\hline
				      & T=100  &T=1000 & T=10000\\ 
				[0.5ex]
				\hline\hline
				VI               & 0.1258 & 1.41      & 12.28    \\
				SWS               & 0.045& 0.056     & 0.079        \\
			 \hline
			\end{tabular}}
		\end{center}
		\vspace{-0.2in}
	\end{table}
\fi

\bibliographystyle{IEEEtran}
\bibliography{IEEEabrv,AgeInfo,ener_harv}

\end{document}